\documentclass[epj]{webofc}
\usepackage[varg]{txfonts}   
\usepackage{slashed}

\wocname{EPJ Web of Conferences}
\woctitle{CONF12}
\newcommand{\be}{\begin{eqnarray}}
\newcommand{\ee}{\end{eqnarray}}

\newcommand{\nn}{\nonumber }

\newcommand{\beq}{\begin{equation}}
\newcommand{\eeq}{\end{equation}}
\newcommand{\bea}{\begin{eqnarray}}
\newcommand{\eea}{\end{eqnarray}}

\begin{document}
\selectlanguage{english}
\title{QCD-inspired determination of NJL model parameters}

\author{Paul Springer\inst{1}\fnsep\thanks{\email{paul.springer@mytum.de}} \and
        Jens Braun\inst{2,3} \and
        Stefan Rechenberger\inst{4}\and
        Fabian Rennecke\inst{5}
}

\institute{Physik Department, Technische Universit\"at M\"unchen, 85747 Garching, Germany 
\and
           Institut f\"ur Kernphysik (Theoriezentrum), Technische Universit\"at Darmstadt, 64289 Darmstadt, Germany
\and
           ExtreMe Matter Institute EMMI, GSI, Planckstra\ss e 1, 64291 Darmstadt, Germany
\and
		   Institut f\"ur Theoretische Physik, Johann Wolfgang Goethe-Universit\"at,
Max-von-Laue-Str.\ 1, D-60438 Frankfurt am Main, Germany
\and
           Institut f\"ur Theoretische Physik, Universit\"at Heidelberg, Philosophenweg 16, 69120 Heidelberg, Germany
}

\abstract{The QCD phase diagram at finite temperature and density has attracted considerable interest
over many decades now, not least because of its relevance for a better understanding of heavy-ion collision experiments. 
Models provide some insight into the QCD phase structure but usually rely on various parameters. 
Based on renormalization group arguments, we discuss how the parameters
of QCD low-energy models can be determined from the fundamental theory
of the strong interaction. We particularly focus
on a determination of the temperature dependence of these parameters 
in this work and comment on the
effect of a finite quark chemical potential. We present first results and argue that our findings can be used to improve the predictive power of 
future model calculations.
}
\maketitle
\section{Introduction}\label{sec:Introduction}
Studies of the chiral phase structure of Quantum Chromodynamics (QCD) at finite temperature~$T$ 
and finite chemical potential~$\mu$ is currently a topic of very active research and is of great importance for many areas of modern physics 
ranging from cosmology over astrophysics to heavy-ion collision experiments. 
Low-energy QCD models, such as Nambu--Jona-Lasinio-type (NJL) and quark-meson-type (QM) models, 
open valuable insights into the dynamics underlying the structure of the QCD phase diagram,
see, e.g., Refs.~\cite{Klevansky:1992qe,Buballa:2003qv,Fukushima:2011jc,Andersen:2014xxa} for reviews.
However, model studies usually require to fix a set of parameters
such that a given set of low-energy observables at vanishing temperature and quark chemical potential is recovered correctly.
The so determined set of parameters is then used to predict, e.g., the location of the chiral phase boundary at finite temperature and/or
quark chemical potential. Unfortunately, different parameter sets may reproduce the correct values of a given set of
low-energy observables at $T=\mu=0$ equally well but may lead to different predictions for the phase diagram, e.g., to different critical temperatures.
In addition, the model parameters may depend on 
the temperature and the quark chemical potential, unless they are fixed at an ultraviolet (UV) scale which is much greater than
the temperatures and chemical potentials under consideration. However, 
the latter is usually not the case. Therefore it is reasonable to assume that model parameters depend on the external
control parameters, such as the temperature or the quark chemical potential.

One of the assumptions underlying the construction of QCD low-energy models is that the dynamics associated with gauge degrees of freedom 
at high-energy scales has been formally integrated out and is absorbed into the parameters of the model. For example, four-quark
interactions are induced by quark-gluon interactions (e.g. two-gluon exchange diagrams) in the high-energy limit. A determination
of the dependence of model parameters on the temperature and the quark chemical potential therefore requires 
a study of quark-gluon dynamics at momentum scales above the UV cutoff scale defining 
the model under consideration. Our guiding principle is to use renormalization group (RG) flow equations
to calculate the~$T$- and~$\mu$-dependence of the parameters of NJL/QM-type models. In this work,
we particularly focus on the computation of gluon-induced quark self-interactions at finite temperature and quark chemical potential which can then be
used to compute the~$T$- and~$\mu$-dependence of the effective quark self-interactions entering NJL/QM-type models. 
More specifically, at some particular scale to be defined below, we shall project the results from our QCD RG-flow calculations 
onto a NJL/QM-type model.

This paper is organized as follows: In Sec.~\ref{sec:FlowEquationsQCD}, we discuss the RG flow of gluon-induced
quark self-interactions. The initial conditions for these flows are set by the standard QCD action.
We analyze the fixed-point structure of the gluon-induced quark self-interactions at finite temperature and quark chemical
potential following Refs.~\cite{Braun:2005uj,Braun:2006jd}, see Ref.~\cite{Braun:2011pp} for a review.
We then discuss the projection of the gluon-induced four-quark interaction channels onto a NJL/QM-type model.
This is detailed in Sec.~\ref{sec:NJLFromQCD} where also the implications of such a parameter determination 
for studies of the phase diagram within this class of low-energy models is discussed. Our summary can be found in~Sec.~\ref{sec:summary}.

\section{Gluon-induced quark self-interactions}\label{sec:FlowEquationsQCD}
For our computation of the RG flow equations of gluon-induced quark self-interactions, we employ an RG equation for the quantum effective action~$\Gamma$,
the {\it Wetterich} equation~\cite{Wetterich:1992yh}. The effective action then depends on the RG scale~$k$ (infrared cutoff scale) which determines the
RG `time'~$t=\ln(k/\Lambda)$ with~$\Lambda$ being the UV scale at which we fix the initial value for the strong coupling~$\alpha = g^2/(4\pi)$.
For our discussion in this section, it suffices to consider the following ansatz for the scale-dependent effective action~$\Gamma_k$ in Euclidean spacetime,
see Refs.~\cite{Gies:2005as,Braun:2005uj,Braun:2006jd,Braun:2011pp}:
\be
\Gamma_k &= & \int \text{d}^4 x \, \, \Bigg\{ \bar{\psi}(\text{i} Z_{\psi} \slashed{\partial}+Z_1 \bar{g} \slashed{A} + \text{i} \gamma_0 \mu)\psi + \frac{1}{4}Z_A F_{\mu \nu}^z F^z_{\mu \nu} \nn\\
&& 
\quad +\frac{1}{2} \Big{[} \bar{\lambda}_- (\text{V}\!-\!\text{A}) + \bar{\lambda}_+ (\text{V}\!+\!\text{A}) +  \bar{\lambda}_{\sigma} (\text{S}\!-\!\text{P})
  + \bar{\lambda}_{\text{VA}} [2(\text{V}\!-\!\text{A})^{\text{adj}}+(1/N_{\text{c}})(\text{V}\!-\!\text{A})]  \Big{]} \Bigg\} \,.
\label{eq:GammaQCD}
\ee
Here, $A_{\mu}=A_{\mu}^z \, t^z$ with $t^z$ being the generators of the group SU($N_\text{c}$) in fundamental representation and~$Z_{A}$ is the wave function
renormalization associated with the gauge fields. The inclusion of the ghost sector and a gauge fixing term is tacitly assumed. 

In our studies below,
we shall always restrict ourselves to {\it Landau gauge} which is known to be an RG fixed point~\cite{Ellwanger:1995qf,Litim:1998qi}. 
Moreover, we have set~$Z_{\psi}=1$ for the wave function renormalization of the quark fields in Eq.~\eqref{eq:GammaQCD} 
which implies that the associated anomalous
dimension~$\eta_{\psi}$ is zero. In the {\it Landau} gauge, this has been indeed found to be the case in the chirally symmetric regime of QCD~\cite{Gies:2003dp}, 
at least if we drop any nontrivial momentum dependencies of the four-quark interactions and consider them in the point-like limit,~$\bar{\lambda}_i(|p_j|\ll k)$. 
The ansatz~\eqref{eq:GammaQCD} respects the SU($N_\text{c}$) gauge symmetry and the U($N_{\rm f}$)$_{\rm L}\, \times$ U($N_{\rm f}$)$_{\rm R}$
flavor symmetry, where~$N_{\rm c}$ and~$N_{\rm f}$ 
denotes the number of colors and flavors, respectively, i.e. we drop~U$_{\rm A}(1)$-violating terms and set the current quark masses to zero.
The first two four-quark interaction channels in Eq.~\eqref{eq:GammaQCD} are defined as follows:
\be
(\text{V}-\text{A})= (\bar{\psi}\gamma_{\mu}\psi)^2+(\bar{\psi}\gamma_{\mu} \gamma_5\psi)^2 \text{ ,} \qquad (\text{V}+\text{A})
= (\bar{\psi}\gamma_{\mu}\psi)^2-(\bar{\psi}\gamma_{\mu} \gamma_5\psi)^{2}\,
\label{eq:QCDChannels1}
\ee
with color ($i,j, \ldots$) and flavor ($a,b, \ldots$) indices assumed to be contracted pairwise. These two channels correspond to 
color and flavor singlets. The remaining two channels have a non-trivial color and flavor structure and are given by
\begin{align}
(\text{S}-\text{P})= (\bar{\psi}^a\psi^b)^2-(\bar{\psi}^a\gamma_5\psi^b)^2 \text{ ,} \qquad (\text{V}-\text{A})^{\text{adj}}
= (\bar{\psi}\gamma_{\mu} \, t^z \, \psi)^2 + (\bar{\psi}\gamma_{\mu} \gamma_5 \, t^z \, \psi)^{2}\,,
\label{eq:QCDChannels2}
\end{align}
where, e.g., $(\bar{\psi}^a\psi^b)^2 \equiv \bar{\psi}^a_i\,\psi^b_i \,\bar{\psi}^b_j\,\psi^a_j$. Any other point-like four-quark interaction channel 
respecting the SU($N_\text{c}$) gauge symmtry and the U($N_{\rm f}$)$_{\rm L}\, \times$ U($N_{\rm f}$)$_{\rm R}$ flavor symmetry
is reducible by means of {\it Fierz} transformations, i.e. our ansatz~\eqref{eq:GammaQCD} is {\it Fierz}-complete in this respect. It is possible to construct
other four-quark interaction channels from these channels which correspond to the ones underlying conventional NJL/QM-type model studies. For example, for~$N_{\rm f}=2$,
the~$(\text{S}-\text{P})$ and~$(\text{V}+\text{A})$ channel can be combined to yield the conventional four-quark channel~$(\text{S}-\text{P})^{\prime}$ associated with
$\sigma$-meson and pion interactions in NJL/QM-type models:
\be
\bar{\lambda}_{\sigma\pi}(\text{S}-\text{P})^{\prime}
=\bar{\lambda}_{\sigma\pi}[(\bar{\psi}\psi)^2-(\bar{\psi} \gamma_5 \vec{\tau} \psi)^2] 
- \bar{\lambda}_{\sigma\pi}[ \text{det}_{\text{f}} \bar{\psi} (1+\gamma_5) \psi + \text{det}_{\text{f}} \bar{\psi} (1-\gamma_5) \psi]\,.
\label{eq:modelchannel}
\ee
Here, the~$\tau_i$'s denote the {\it Pauli} matrices and the determinant is taken in flavor space. In terms of the four-quark couplings, 
this channel corresponds to the the combination~$\bar{\lambda}_{\sigma\pi}=\bar{\lambda}_{\sigma} - \frac{2}{3} \bar{\lambda}_{+} $. The first
term in Eq.~\eqref{eq:modelchannel} is the conventional scalar-pseudoscalar channel associated with $\sigma$-meson and pion interactions in model studies.
The second term in Eq.~\eqref{eq:modelchannel} has the same structure as a term associated with topologically non-trivial gauge configurations 
that break the U$_{\rm A}(1)$ symmetry~\cite{tHooft:1976fv}. As we do not take into account a breaking of the U$_{\rm A}(1)$ symmetry, the first and the second
term in Eq.~\eqref{eq:modelchannel} contribute with the same strength~$\bar{\lambda}_{\sigma\pi}$. Note that we are only interested in a computation
of the RG flows of these potentially gluon-induced four-quark interactions at intermediate and high momentum scales above the chiral symmetry
breaking scale. Therefore this approximation still appears to be reasonable as a strong deviation between the two couplings is only expected to emerge 
in the infrared (IR) regime close and below the chiral symmetry breaking scale~\cite{Mitter:2014wpa}. Below, we shall therefore 
identify the coupling~$\bar{\lambda}_{\sigma\pi}$ with the four-quark coupling appearing in conventional NJL/QM-type studies.
\begin{figure}
\centering
\includegraphics[width=0.8\textwidth]{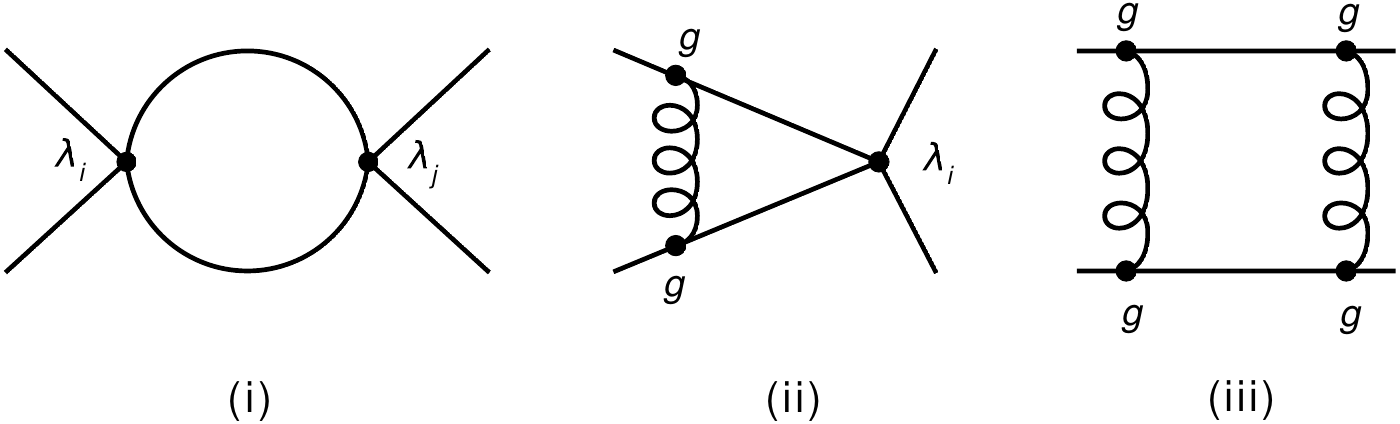}
\caption{\small{The Feynman diagrams (i), (ii), and (iii) are associated with the coefficients~$A_{jk}^{(i)}$, $B_j^{(i)}$, and $C^{(i)}$ in the
flow equations~\eqref{eq:ffflow} of the four-quark couplings, respectively.}}
\label{fig:feynman}
\end{figure}

Using the {\it Wetterich} equation together with the ansatz~\eqref{eq:GammaQCD}, we obtain the
flow equations for the four-quark interactions which can be conveniently 
written in the following general form in the point-like limit~\cite{Gies:2005as,Braun:2005uj,Braun:2006jd,Braun:2011pp}:
\be
\partial_t \lambda_{i} = 2 \lambda_{i} - \sum_{j,k}A_{jk}^{(i)} \lambda_{j}\lambda_{k} - 
\sum_j B_j^{(i)} \lambda_{j}g^2 - C^{(i)} g^4\,,
\label{eq:ffflow}
\ee
where $i\in\{-,+,\sigma,\text{VA}\}$. The renormalized dimensionless couplings are defined as $\lambda_i = k^2\bar{\lambda}_i$ and~$g^2 = \bar{g}^2Z_1^2/Z_A$ 
with~$\bar{g}$ being the bare gauge coupling. Recall that~$Z_{\psi}\equiv 1$ in our present study. The coefficients~$A_{jk}^{(i)}$, $B_j^{(i)}$, and $C^{(i)}$ 
are associated with loop integrals, see Fig.~\ref{fig:feynman}, and depend on the number of flavors, colors, the dimensionless temperature~$T/k$ and, in general, also on
the dimensionless quark chemical potential~$\mu/k$. In the limit~$T\to 0$ and~$\mu\to 0$, these coefficients become regularization-scheme
dependent constants. For our explicit numerical studies, we employ the so-called 4$d$ 
exponential regualor~\cite{Jungnickel:1995fp} for which the exact form of the coefficients~$A_{jk}^{(i)}$, $B_j^{(i)}$, and $C^{(i)}$ as a function of~$T/k$
can be found in Ref.~\cite{Braun:2006jd}. For other regulator functions (e.g. linear regulators~\cite{Litim:2000ci,Litim:2006ag}), 
the general form of the flow equations~\eqref{eq:ffflow} remains unchanged, only
the quantitative dependence of the coefficients~$A_{jk}^{(i)}$, $B_j^{(i)}$, and $C^{(i)}$ on, e.g., the temperature changes. As initial condition for the four-quark
couplings, we choose~$\lambda_i\to 0$ for~$k\to \Lambda$, which ensures that our ansatz~$\Gamma_k$ for the effective
action is identical to the QCD action in this limit, i.e.~$\Gamma_{k\to\Lambda}\to S_{\rm QCD}$.
\begin{figure}
\centering
\sidecaption
\includegraphics[width=0.45\textwidth]{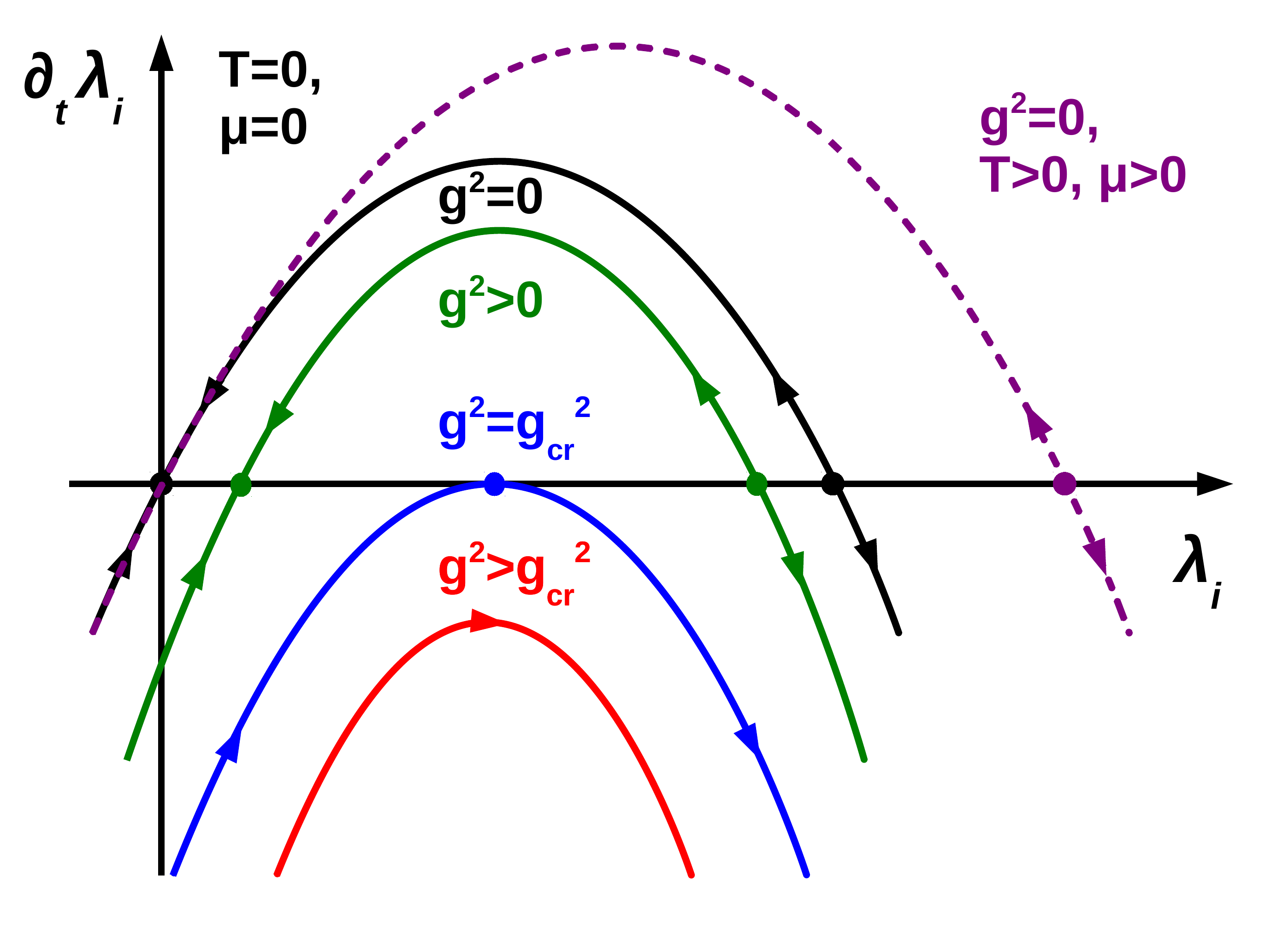}
\caption{\small{Sketch of the $\beta$-function~$\eqref{eq:ffflow}$ of a four-quark coupling~$\lambda_i$ 
evaluated for different values of the strong coupling $g^2$. 
For a sufficiently strong gauge coupling $g^2>g_{\rm cr}^2$, the RG flow of the four-quark couplings is no longer governed by the presence
of real-valued fixed points but increase rapidly towards the infrared, thereby indicating the onset of spontaneous
chiral symmetry breaking.}}
\label{fig:FlowsSketch}
\end{figure}

The flow equations of the couplings $\lambda_i$ provide us with a simple picture of the chiral dynamics in gauge theories~\cite{Gies:2003dp, Gies:2005as,Braun:2006jd,Braun:2011pp}.
To see this, we consider the flow equations~\eqref{eq:ffflow} and Fig.~\ref{fig:FlowsSketch}. From Eq.~\eqref{eq:ffflow}, it follows
that the $\beta$-function associated with the coupling~$\lambda_i$ is an inverted parabola as a function of~$\lambda_i$.\footnote{Here, we assume that 
the~$A_{jj}^{(i)}$'s are positive~\cite{Gies:2005as,Braun:2005uj,Braun:2006jd,Braun:2011pp}.}  
In the limit~$T\to0$, $\mu\to 0$, and~$g^2\to 0$, the $\beta$-functions of the four-quark couplings have two fixed points: an 
IR attractive Gau\ss ian and an IR repulsive non-trivial fixed point, see 
black solid line in Fig.~\ref{fig:FlowsSketch}.\footnote{The arrows in Fig.~\ref{fig:FlowsSketch} indicate the direction of the 
RG flow towards the IR limit.} Since we choose~$\lambda_i=0$ as initial condition for all four-quark
couplings, the system stays at the Gau\ss ian fixed point, i.e. remains non-interacting on all scales in this limit.
For increasing gauge coupling~$g^2$ but still for~$T=0$ and~$\mu=0$, the two fixed points (i.e. zeroes) of the $\beta$-functions of the couplings~$\lambda_i$ 
approach each other, see Fig.~\ref{fig:FlowsSketch}. Increasing the gauge coupling further, we observe that
the two fixed points annihilate each other at a critical value of the gauge coupling~$g^2=g^2_{\rm cr}$. Increasing now
the gauge coupling beyond this critical value, the parabola is pushed below the~$\lambda_i$ axis and 
the flow is no longer governed by the fixed points and the four-quark couplings start to increase rapidly and 
approach a divergence at a finite scale~$k=k_{{\rm SB}}$, indicating the onset of spontaneous 
symmetry breaking associated with the formation of corresponding condensates. In fact, the couplings~$\lambda_i$
are inverse proportional to the mass parameter of a {\it Ginzburg}-{\it Landau}-type effective potential in a bosonic
formulation, see Ref.~\cite{Braun:2011pp} for details and also our discussion in Sec.~\ref{sec:NJLFromQCD} below.
At this point, we have traced the question of chiral symmetry breaking back to the strength of the coupling~$g$
relative to the critical coupling~$g_{\rm cr}$.

For finite temperatures~\cite{Gies:2005as,Braun:2006jd,Braun:2011pp}, the Feynman diagrams contributing to the RG
flow of the four-quark couplings are parametrically suppressed as all quark modes acquire thermal masses. In other words,
the coefficients~$A_{jk}^{(i)}$, $B_j^{(i)}$, and $C^{(i)}$ tend to zero for~$T\to\infty$. In particular, we have~$A_{jk}^{(i)}\to 0$
for~$T\to \infty$. As a consequence, the parabolas associated with the $\beta$-functions of the four-quark couplings become broader 
and higher for increasing temperature, see Fig.~\ref{fig:FlowsSketch}. Thus, the critical value of the gauge coupling increases with increasing~$T/k$. 
This observation already suggests that a critical value of the temperature exists above which the four-quark couplings remain
finite on all scales, i.e. the system remains in the chirally symmetric phase. In Refs.~\cite{Gies:2005as,Braun:2006jd}, this
interplay of the fixed points of gluon-induced four-quark interactions and the running gauge coupling has been used to obtain
a first-principles estimate for the chiral phase transition temperature as a function of the number of quark flavors 
which has indeed been found to agree very well with state-of-the-art lattice QCD results~\cite{Aoki:2009sc,Cheng:2009zi}.
Moreover, a corresponding fixed-point analysis in the presence of a finite magnetic field provides an explanation of inverse magnetic 
catalysis in QCD~\cite{Braun:2014fua,Mueller:2015fka}.

At finite quark chemical potential, the situation is similar to the case of finite temperature. In particular,
the coefficients~$A_{jk}^{(i)}$ associated with the pure quark loop in Fig.~\ref{fig:feynman} tend to zero for~$\mu/k \to \infty$ as well.
Thus, the parabola becomes again broader and higher when~$\mu/k \to \infty$ is increased, see also Fig.~\ref{fig:FlowsSketch}. This 
suggests that the critical value of the gauge coupling also increases with increasing~$\mu/k$, implying the existence of a critical value
of the quark chemical potential above which no spontaneous symmetry breaking occurs. 

For an actual numerical solution of the flow equations of the four-quark couplings, we need to specify the running of the strong coupling~$g^2$.
In this work, we do not compute the running coupling but use it as an input. More specifically, we employ the running coupling from Refs.~\cite{Braun:2005uj,Braun:2006jd} 
where it has been computed within the background-field formalism at zero and finite temperature 
with the same exponential regulator function as we use here. With respect to an analysis of the effect of a finite quark
chemical potential, we therefore restrict ourselves to a qualitative discussion. In any case, a word of caution needs to be added here: At high momentum scales (i.e.
in the perturbative limit), the running of the coupling~$g^2$ computed within the background-field formalism is indeed identical to the one of the 
coupling~$g_{\bar{\psi}\slashed{A}\psi}^2=\bar{g}Z_1^2/(Z_{\psi}^2Z_A)$ associated
with the quark-gluon vertex which essentially enters our RG flow equations of the four-quark couplings. At intermediate or even low momentum scales, however,
these two definitions of the strong coupling do not necessarily agree. 
For example, at zero temperature and intermediate momentum scales~$k\sim {\mathcal O}(1\,\text{GeV})$~\cite{Mitter:2014wpa,Braun:2014ata},
the coupling~$g_{\bar{\psi}\slashed{A}\psi}^2$ has been found to be greater than the background-field
running coupling. Therefore, the use of the background-field coupling from Refs.~\cite{Braun:2005uj,Braun:2006jd}
in this work may be considered as an approximation.
A detailed analysis of this issue will be presented elsewhere. Here, we only state that the scale~$\Lambda_{\rm I}$ 
(see Eq.~\eqref{eq:Lambda0Fix} below) at which we project the gluon-induced four-quark couplings 
onto a QCD low-energy model
increases already at zero temperature when we employ~$g_{\bar{\psi}\slashed{A}\psi}^2$ in our calculations. 
In any case, we shall fix the coupling in the deep UV regime and use~$\alpha(k=20\,\text{GeV})\simeq 0.163$~\cite{Bethke:2002rv}.

In the following we do not aim at a direct computation of the chiral finite-temperature phase boundary from an analysis of gluon-induced four quark-interactions 
as detailed in Refs.~\cite{Braun:2005uj,Braun:2006jd}. We rather employ the set of flow equations~\eqref{eq:ffflow} to compute the temperature dependence
of the four-quark couplings entering NJL/QM-type model studies. Therefore, we do not search for a divergence of the gluon-induced four-quark couplings
but evaluate them already at a higher scale~$\Lambda_{\rm I} > k_{\rm SB}$ to read off
the value of the temperature-dependent four-quark couplings and use them as input for QCD low-energy model studies. Thus, the scale~$\Lambda_{\rm I}$ is identified
with the UV cutoff scale of the low-energy models. Still, an analysis of the fixed-point structure of the four-quark couplings is very useful to obtain
an analytic understanding of the strength of the gluon-induced four-quark couplings which are eventually used to fix the parameters of QCD low-energy models. 
From Fig.~\ref{fig:FlowsSketch}, we deduce that an IR attractive Gau\ss ian fixed point is approached at high momentum scales as 
the gauge coupling tends to zero logarithmically in this regime. 
Towards the IR regime the gauge coupling increases and the IR attractive Gau\ss ian fixed point
becomes an interacting IR attractive fixed point. The four-quark couplings follow this fixed point. At least in the weak-coupling limit,  the 
fixed-point values~$\lambda_i^{\ast}$ of the four-quark couplings can be computed analytically. Up to numerical factors, we find
\be
\lambda_i^{\ast}(T,k) \sim C^{(i)}g^4(T,k)\,,
\label{eq:lambdagi}
\ee
where~$\lambda_i^{\ast}$ and~$g$ depend on both the temperature and the RG scale~$k$. Thus, the four-quark couplings are not parameters 
of our study but indeed induced by the fundamental quark-gluon interactions. Our choice to set the four-quark couplings to zero at the initial high-momentum
scale is consistent with the appearance of the Gau\ss ian fixed point in this limit. 
From Eq.~\eqref{eq:lambdagi}, we obtain immediately a first estimate of the temperature-dependence of the effective four-quark couplings~$\lambda_{\text{NJL},i}$ entering
QCD low-energy model studies:
\be
\lambda_{\text{NJL},i}(T,\Lambda_0) \sim C^{(i)}g^4(T,\Lambda_0)\,.\label{eq:ffana}
\ee
Here, the scale~$\Lambda_0$ is the UV cutoff scale of the low-energy model under consideration which we assume to be kept fixed also at finite temperature. 
Recall that the~$C^{(i)}$'s are associated with the two-gluon exchange diagram in Fig.~\ref{fig:feynman} and depend in general also on the temperature. In any case, 
the temperature dependence of the strong coupling directly affects the temperature dependence of the parameters of the low-energy model. We conclude
this section by noting that the arguments leading to Eq.~\eqref{eq:ffana} are very general and can be straightforwardly extended to external parameters other
than the temperature, such as a quark chemical potential, isospin chemical potential, and a magnetic field.

\section{Low-energy model from gluon-induced quark self-interactions}\label{sec:NJLFromQCD}
For the low-energy sector of QCD, we now consider the standard NJL model with two massless quark flavors and~$N_{\rm c}=3$ 
colors~\cite{Klevansky:1992qe} defined by the following ansatz for the so-called classical action:
\be
S_{\text{NJL}} = \int \text{d}^4\text{x} \, \, \left\{ \bar{\psi}(\text{i} \slashed{\partial}+\text{i}\gamma_0 \mu)\psi
 +\frac{1}{2}{\bar{\lambda}_{\text{NJL}}} [(\bar{\psi}\psi)^2-(\bar{\psi} \gamma_5 \vec{\tau} \psi)^2] \right\}\,.
\label{eq:QCDModel}
\ee
As we shall see below, this model comes with three parameters. As it is non-renormalizable, we define it with an UV cutoff~$\Lambda_0$ which
represents one of these parameters. As a consequence, the regularization scheme belongs to the definition of the model. Here, we shall 
use the same exponential regulator function as in our study of the RG flow of the gluon-induced four-quark interactions in order to facilitate 
the projection of the latter on the NJL model. In our ansatz~\eqref{eq:QCDModel} we have dropped the 't~Hooft term which appears
in the definition of our four-quark channel associated with the coupling~$\lambda_{\sigma\pi}$, see Eq.~\eqref{eq:modelchannel}.
In our present exploratory study of 
the temperature dependence of parameters of low-energy models, we shall assume that the gluon-induced~$\lambda_{\sigma\pi}$-channel
can be identified with the~$\lambda_{\text{NJL}}$-channel in the NJL model at its UV cutoff scale. Note that~$\lambda_{\text{NJL}}$ is in general
a scale-dependent quantity.
Two comments are in order here: First, this implies
that we assume that the breaking of the U$_{\text{A}}(1)$ symmetry is still small at least at intermediate and large momentum scales and only
becomes significant in the low-energy limit, which indeed appears reasonable~\cite{Mitter:2014wpa}. Second, as we shall study the NJL
model in the mean-field approximation, we add that this approximation is plagued by a so-called {\it Fierz} ambiguity which cannot be resolved even
if we considered a {\it Fierz}-complete ansatz for the (effective) action~\cite{Jaeckel:2002rm}. Moreover, our NJL model ansatz is
not even {\it Fierz}-complete. Therefore the projection of our gluon-induced four-quark interactions onto the NJL model~\eqref{eq:QCDModel} is 
to some extent ambiguous and we shall not aim at quantitative studies but rather aim to provide qualitative guidance on how QCD
low-energy models can be amended on the basis of the fundamental theory.
\begin{figure}
\begin{center}
\includegraphics[width=0.495\textwidth]{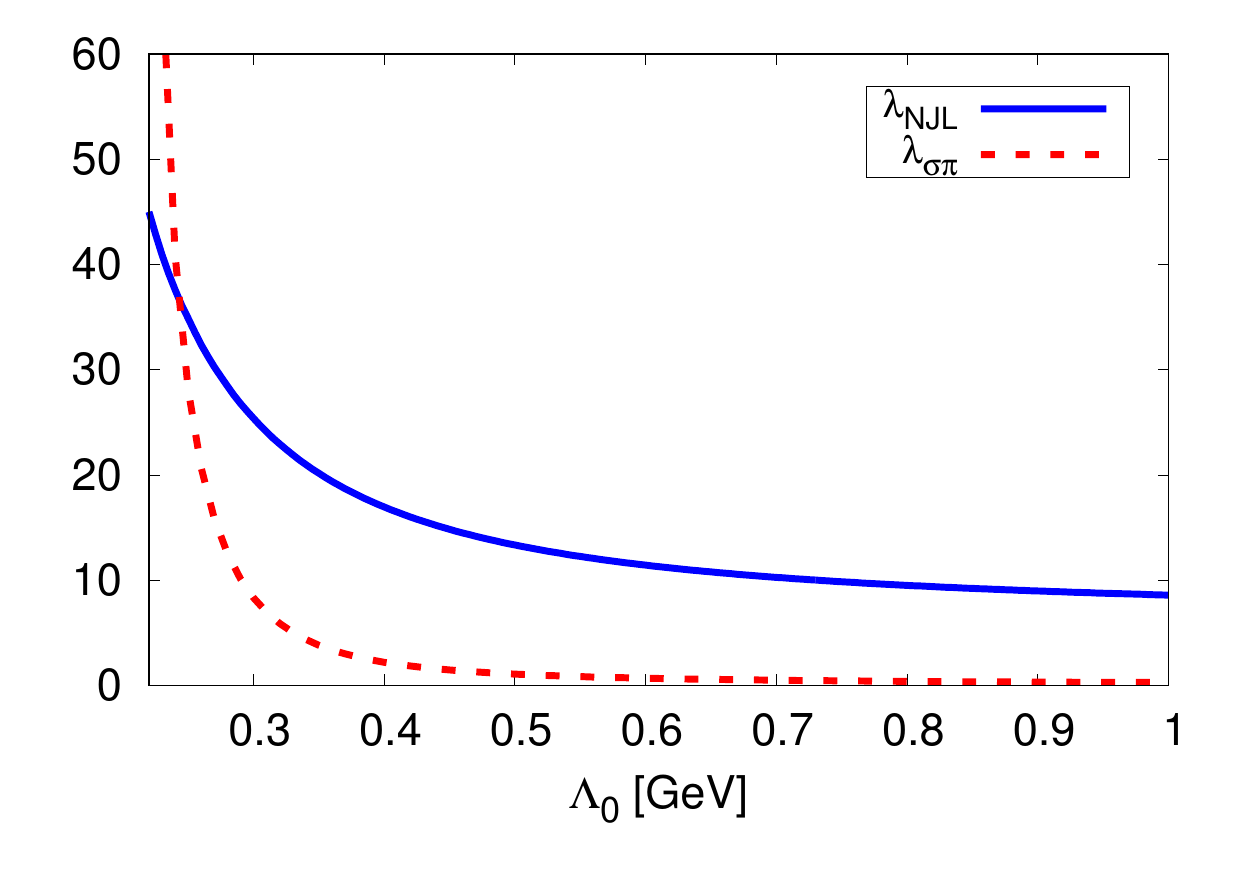}
\includegraphics[width=0.495\textwidth]{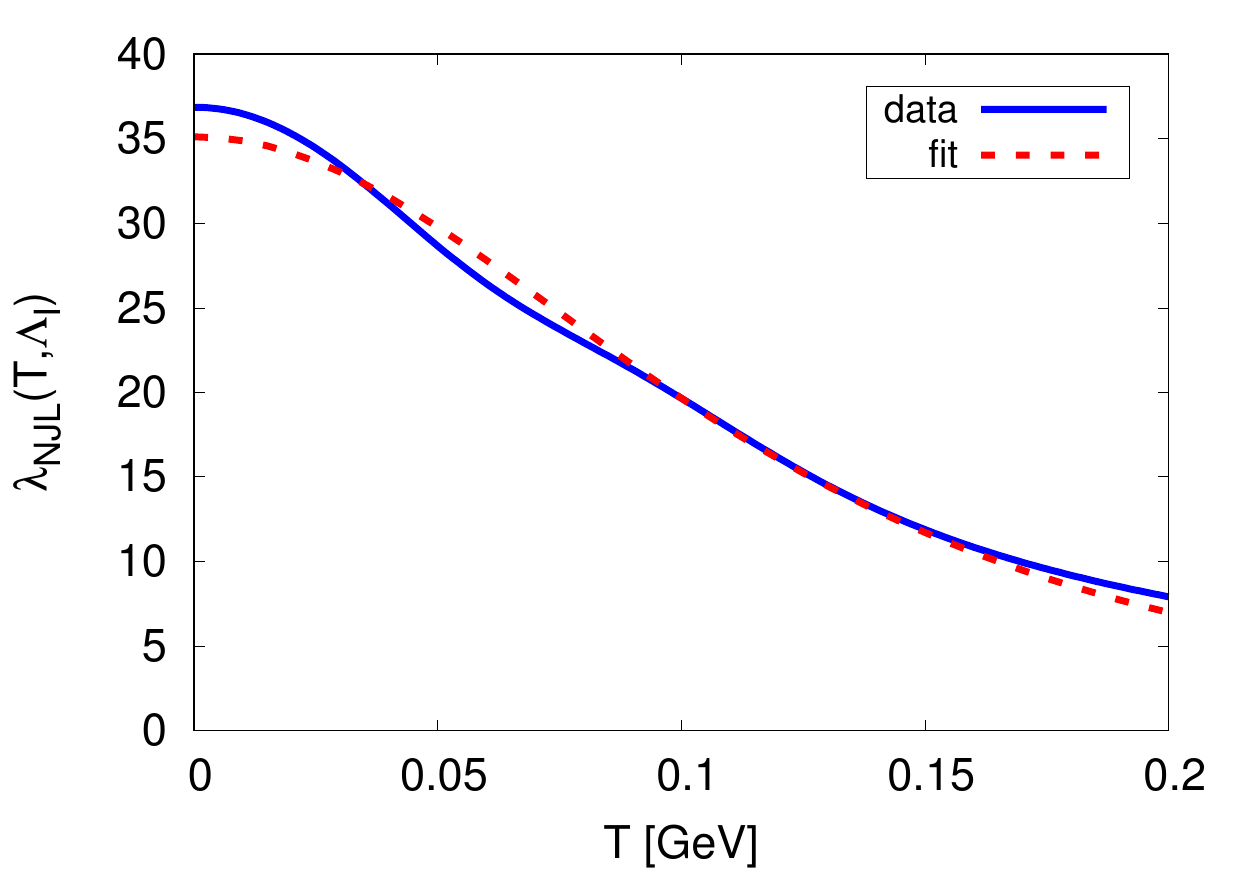}
\end{center}
\caption{Left panel: Dimensionless four-quark coupling~$\lambda_{\text{NJL}}=\Lambda_0^2\bar{\lambda}_{\text{NJL}}$ 
of the NJL model as a function of the UV cutoff~$\Lambda_0$ for a fixed 
quark mass~$\bar{m}_q\approx 0.3\,\text{GeV}$ versus the  
gluon-induced four-quark coupling~$\lambda_{\sigma\pi}$ evaluated
at~$k=\Lambda_0$. Right panel: Temperature dependence of the NJL-model 
coupling~$\lambda_{\text{NJL}}(T,\Lambda_{\rm I})=\Lambda_{\rm I}^2\bar{\lambda}_{\text{NJL}}(T,\Lambda_{\rm I})$ 
at the projection scale~$\Lambda_{\rm I}$ as obtained
from the temperature dependence of gluon-induced four-quark interactions (blue line) compared to a fit of the function~\eqref{eq:fit} to 
the numerical data (red dashed line).}
\label{fig:3}
\end{figure}

For a computation of low-energy observables, the purely fermionic formulation of the NJL model in 
Eq.~\eqref{eq:QCDModel} may not be  the most convenient choice since
it requires to resolve the momentum-dependence of, e.g., the four-quark vertex. Therefore, a partially bosonized
formulation  of  the NJL model might  be  better  suited. Such a formulation of the action can be obtained
straightforwardly from a {\it Hubbard-Stratonovich} transformation of the underlying path integral:
\be
S_{\text{PB}} = \int \text{d}^4\text{x} \left\{ \bar{\psi}(\text{i} \slashed{\partial}+\text{i}\gamma_0 \mu)\psi
  + \text{i} \bar{h} \bar{\psi} (\sigma + \text{i} \vec{\tau} \cdot \vec{\pi} \gamma_5) \psi + \frac{1}{2} \bar{m}^2 (\sigma^2 + \vec{\pi}^{\,2})\right\}\,,
\label{eq:QCDModelBosonized}
\ee
where~$\sigma \sim \bar{\psi}\psi$, $\vec{\pi} \sim \bar{\psi}\gamma_5\vec{\tau}\psi$, and $\bar{h}$ is a real valued {\it Yukawa} coupling. At
the bosonization scale (i.e. UV cutoff scale), the four-quark coupling~$\bar{\lambda}_{\text{NJL}}$ can be related straightforwardly to the {\it Yukawa} coupling and 
the bosonic mass parameter~$\bar{m}^2$. We have $\bar{\lambda}_{\text{NJL}} =  \frac{\bar{h}^2}{\bar{m}^2}$. Thus,
a diverging four-quark coupling is associated with vanishing mass parameter~$\bar{m}^2$, i.e. the onset of spontaneous chiral 
symmetry breaking, see Ref.~\cite{Braun:2011pp} for details. Note that, including the UV cutoff scale~$\Lambda_0$, our model
depends on three parameters rather than two as suggested by Eq.~\eqref{eq:QCDModel}, namely~$\bar{m}^2$,~$\bar{h}$, and~$\Lambda_0$.
The {\it Yukawa} coupling is indeed
marginally relevant and is used to adjust the constituent quark mass, see Refs.~\cite{Braun:2011pp,Braun:2012zq} for a detailed analysis
of the fixed-point structure of this class of models.

From the stationary condition for the effective action associated with the action~$S_{\text{PB}}$,
we find the gap equation in the mean-field approximation 
which determines the constituent quark mass~$\bar{m}_q=\bar{h}\sigma_0$:
\be
0= \frac{\bar{m}_q^2}{2\bar{\lambda}_{\text{NJL}}} + \int_0^{\Lambda_0} {\rm d}k\, f_{k}(T,\mu,\bar{m}_q)\,,
\label{eq:gap}
\ee
where the function~$f_k$ depends on the $T$- and~$\mu$-dependent quark propagator and the regularization scheme. For details on the 
computation of this gap equation for general regulators, we refer the reader to Ref.~\cite{Braun:2011pp}. Here, we only
state that the integral over the RG scale~$k$ in Eq.~\eqref{eq:gap} can be directly related to the momentum integration 
in case of a sharp momentum cutoff.

In general, the parameters of NJL/QM-type models are fixed in the limit~$T=0$ and~$\mu=0$.
Fixing the constituent quark mass~$\bar{m}_q$ to a specific value in this limit, i.e.~$\bar{m}_q \approx 0.3\,\text{GeV}$ in 
our present study, the gap equation~\eqref{eq:gap} can be solved for~$\bar{\lambda}_{\text{NJL}}$, yielding the strength
of the effective four-quark coupling~$\bar{\lambda}_{\text{NJL}}$ at~$T=0$ and~$\mu=0$ as a function of~$\Lambda_0$, 
$\bar{\lambda}_{\text{NJL}}\equiv \bar{\lambda}_{\text{NJL}}(\Lambda_0)$. 

In the left panel of Fig.~\ref{fig:3}, we 
show~$\Lambda_0^2\bar{\lambda}_{\text{NJL}}$ as a function of~$\Lambda_0$ which
remains finite in the limit~$\Lambda_0\to \infty$:
\mbox{$\lim_{\Lambda_0\to\infty}\Lambda_0^2\bar{\lambda}_{\text{NJL}}(\Lambda_0)\simeq 6.58$}. On the other hand, the gluon-induced 
four-quark interactions considered in Sec.~\ref{sec:FlowEquationsQCD} tend to zero at large momentum scales.
In order to project the gluon-induced four-quark interactions onto our low-energy model, we determine a projection scale~$\Lambda_{\rm I}$ at~$T=0$
and~$\mu=0$ which is defined to be the intersection point of~$\bar{\lambda}_{\text{NJL}}$ and the corresponding 
gluon-induced four-quark interaction~$\bar{\lambda}_{\sigma\pi}$:
\be
\bar{\lambda}_{\text{NJL}}(\Lambda_{\rm I}) \stackrel{!}{=} \bar{\lambda}_{\sigma\pi}(\Lambda_{\rm I})\,.\label{eq:Lambda0Fix}
\ee
By construction, the intersection point~$\Lambda_{\rm I}$ is also the UV cutoff scale of our QCD low-energy model in the following.
Since we determine~$\bar{\lambda}_{\text{NJL}}$
by fixing the constituent quark mass to a specific value, the {\it Yukawa} coupling is fixed to a specific value in our mean-field study 
and the relation~\eqref{eq:Lambda0Fix}
therefore corresponds to a determination of the bosonic mass parameter~$\bar{m}^2=\bar{h}^2/\bar{\lambda}_{\sigma\pi}(k=\Lambda_{\rm I})$ in the 
bosonic formulation~\eqref{eq:QCDModelBosonized} of our low-energy model.
In the left panel of Fig.~\ref{fig:3}, we 
show~${\lambda}_{\sigma\pi}(k=\Lambda_0)$ (red dashed line) and~${\lambda}_{\text{NJL}}(\Lambda_0)$ (blue line)
and observe that the two functions indeed have a single intersection point at~$\Lambda_{\rm I}\approx 0.243\,\text{GeV}$. We add that 
the so determined UV scale~$\Lambda_{\rm I}$ appears to be small compared to the typical 
values of the UV cutoff scale in model studies. However, recall that a naive comparison between
these values is potentially misleading as this scale does not represent a physical observable but depends on the 
employed regularization scheme. From a phenomenological point of view, 
the scale~$\Lambda_{\rm I}$ agrees in our case almost identically with the chiral symmetry breaking scale which sets the scale for 
all low-energy observables. Most importantly, we add that the actual value of this scale is sensitive to the running of the
strong coupling at momentum scales~$\sim {\mathcal O}(0.5 - 1\,\text{GeV})$. Recall that~$\lambda_{\sigma\pi} \sim g^4$. 
A detailed analysis of this observation is left to future work. 

We note that the slopes of~$\lambda_{\text{NJL}}$ and~$\lambda_{\sigma\pi}$ as a function of~$\Lambda_0$ are different,
in particular at the intersection point~$\Lambda_{\rm I}$, see left panel of Fig.~\ref{fig:3}. Thus, the transition
between the QCD RG-flows at high momentum scales and the QCD low-energy model is not smooth. A smooth transition between
quarks and gluons on the one hand and hadronic degrees of freedom on the other hand without any fine-tuning requires
more elaborate techniques, such a dynamical hadronization techniques in the RG 
context~\cite{Gies:2002hq,Pawlowski:2005xe,Braun:2008pi,Floerchinger:2009uf,Mitter:2014wpa,Braun:2014ata}. Still,
the projection scale~$\Lambda_{\rm I}$ has also been estimated to be of the order of the constituent quark mass when
these techniques are employed~\cite{Rennecke:PhD}. In any case,
once we have determined the scale~$\Lambda_{\rm I}$, we keep it fixed in our finite-temperature studies. The temperature-dependence
of the NJL model coupling~$\lambda_{\text{NJL}}$ is then obtained from an evaluation of the temperature-dependent 
gluon-induced four-quark coupling~$\lambda_{\sigma\pi}$ at the scale~$\Lambda_{\rm I}$: $\lambda_{\text{NJL}}(T)=\lambda_{\sigma\pi}(T,\Lambda_{\rm I})$.
In the right panel of Fig.~\ref{fig:3} we show our results for the temperature-dependence of the dimensionless NJL model 
coupling~$\lambda_{\text{NJL}}(T)=\Lambda_{\rm I}^2\bar{\lambda}_{\text{NJL}}(T)$. We observe that~$\lambda_{\text{NJL}}(T)$ decreases
with increasing temperature. The actual functional dependence of~$\lambda_{\text{NJL}}(T)$ on the temperature can be well described by
the following parametrization:
\be
\lambda_{\text{NJL}}(T)\approx \frac{\tilde{\lambda}_{\text{NJL}}}{\left[\ln \left( \frac{\tilde{\Lambda}_{\rm I}^2 + T^2}{\Lambda_{\text{QCD}}^2}\right)\right]^2}\,,
\label{eq:fit}
\ee
which is inspired from our analytic considerations, see Eq.~\eqref{eq:ffana}, and the one-loop running of the strong 
coupling~$g^2(q^2) \sim 1/\ln(q^2/\Lambda_{\text{QCD}}^2)$. In Eq.~\eqref{eq:fit},~$\tilde{\Lambda}_{\rm I}$ and~$\tilde{\lambda}_{\text{NJL}}$ are  
parameters which we have estimated from a fit to the numerical data, and~$\Lambda_{\text{QCD}}\approx 0.371\,\text{GeV}$ is the position
of the {\it Landau} pole of the running coupling in the one-loop approximation. We find~$\tilde{\lambda}_{\text{NJL}} \approx 1.09$
and~$\tilde{\Lambda}_{\rm I}\approx 0.405\,\text{GeV}$. The scale~$\tilde{\Lambda}_{\rm I}$ is essentially related
to the UV cutoff scale~$\Lambda_{\rm I}$ of the low-energy model in the sense that both scales are associated with the UV extent of the model.
\begin{figure}
\sidecaption
\includegraphics[width=0.51\textwidth]{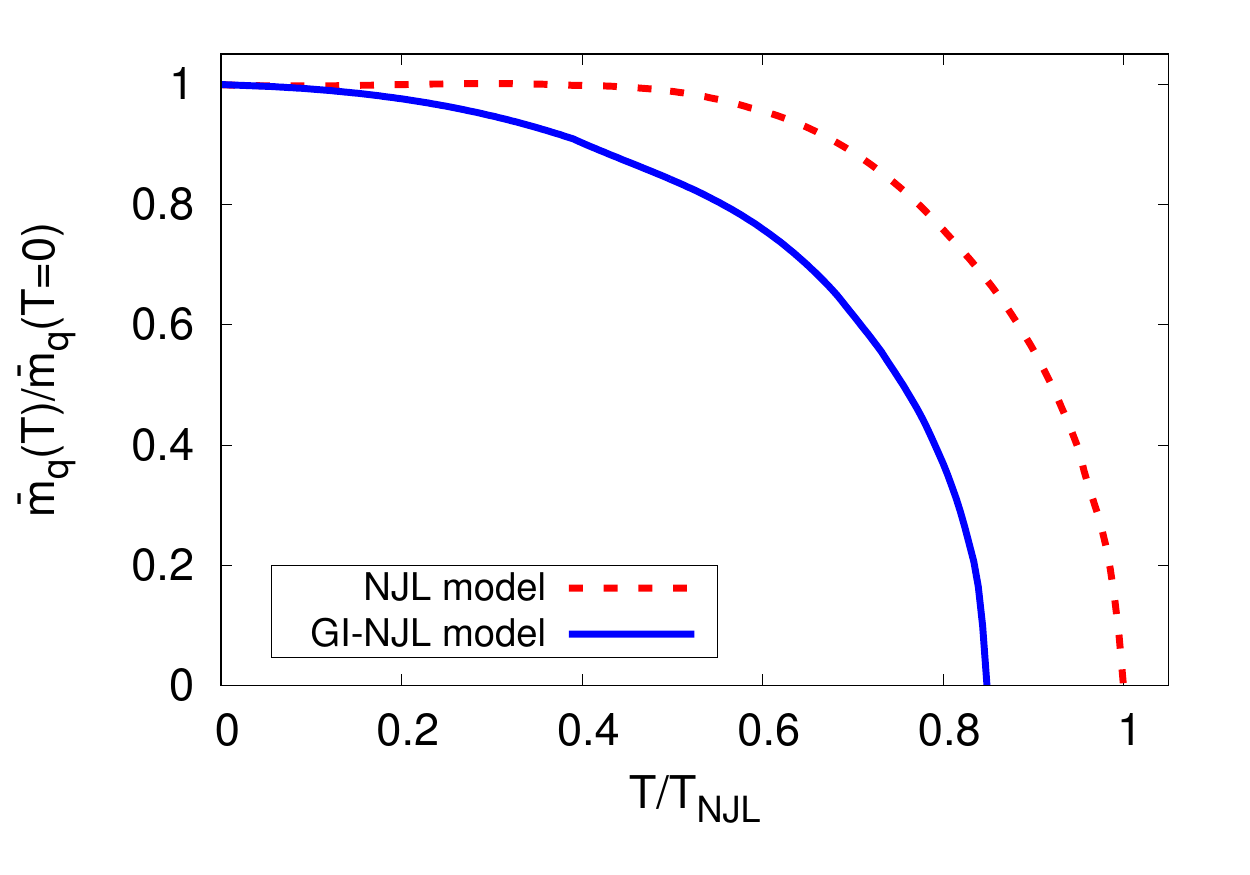}
\caption{Constituent quark mass~$\bar{m}_q$ as a function of the temperature~$T$ as obtained from the standard NJL model and the
GI-NJL model. Here,~$T_{\text{c},\text{NJL}}$ denotes the chiral critical temperature obtained from the NJL model study.}
\label{fig:4}
\end{figure}

The temperature-dependent effective four-quark coupling~$\lambda_{\text{NJL}}(T)=\Lambda_{\rm I}^2\bar{\lambda}_{\text{NJL}}(T)$ can now be used
to study, e.g, the finite-temperature chiral phase transition with our low-energy model. To this end, we insert~$\bar{\lambda}_{\text{NJL}}(T)$ into
the gap equation~\eqref{eq:gap} to obtain the constituent quark mass~$\bar{m}_q$. To distinguish the results for the temperature dependence 
of~$\bar{m}_q$~($\sim$ chiral order parameter) from a conventional NJL model study with a
temperature-independent four-quark coupling, we refer to the low-energy model with a 
temperature-dependent four-quark coupling parameter as gluon-induced NJL (GI-NJL) model.

Since the constituent quark mass decreases with decreasing coupling~$\lambda_{\text{NJL}}$ 
for fixed UV cutoff scale and {\it Yukawa} coupling~$\bar{h}$, we expect that the chiral phase transition temperature obtained
from the GI-NJL model is smaller than the one from a conventional NJL model study. In Fig.~\ref{fig:4}, we show
the constituent quark mass as a function of the temperature where the same value for the UV cutoff scale~$\Lambda_{\rm I}$ has been
used in the NJL as well as the GI-NJL model to ensure comparability.
Indeed, we find 
that the chiral critical temperature is reduced by~$\sim 15\%$ when we employ a temperature-dependent
four-quark coupling~$\lambda_{\text{NJL}}$. From the conventional NJL model, we obtain~$T_{\text{c},\text{NJL}}\approx 0.103\,\text{GeV}$
for the absolute value of the chiral phase transition temperature. Note that the apparent smallness of the absolute value of the critical temperature 
is to some extent related to the actual value of the UV cutoff scale~$\Lambda_{\rm I}$ which is found to be comparatively small in our present exploratory study.
In any case, we observe that the chiral phase transition temperature becomes smaller when we use
the temperature-dependent four-quark coupling. Whereas our present study may not yet be correct on 
a quantitative level, the mechanisms underlying this observation appear to be quite general.
First studies of the chiral phase boundary in the plane spanned by the temperature and the quark chemical
potential  indicate that the phase transition temperature also becomes smaller for a given value of~$\mu$
when we employ a $T$- and~$\mu$-dependent effective four-quark coupling in the QCD low-energy model, although
the effect seems to become smaller with increasing~$\mu$.

\section{Summary}\label{sec:summary}
In this work we explored a possible dependence of QCD low-energy model parameters on external control
parameters, with an emphasis on the temperature dependence of four-quark interactions. In particular, we have shown on very general grounds
how RG flow equations at high momentum scales allow for a first-principles computation of the 
parameters of QCD low-energy models. For the standard NJL model, we find
that the model parameter associated with the four-quark coupling decreases with increasing temperature
which eventually results in a decrease of the chiral phase transition temperature, also at finite quark chemical potential.
More specifically, for~$\mu=0$, we find that the use of the temperature-dependent four-quark coupling lowers the chiral phase
transition temperature by~$\sim 15\%$ compared to the result from a study with a temperature-independent coupling.
Our present study is not yet quantitative. Still,
the mechanisms, namely the thermal suppression of gluon-induced four-quark interactions at high momentum
scales, indeed appears to be very general. In fact, from our analysis of the RG flow equations of the gluon-induced 
four-quark interactions, we have even deduced a simple functional form for the temperature dependence of
the four-quark interaction of low-energy models which may already turn out to be useful  
to guide future model studies of the QCD phase diagram.

{\it Acknowledgments.-} J.B. is grateful to H. Gies for useful discussions.
Moreover, as members of the {\it fQCD collaboration}~\cite{fQCD}, J.B. and F.R. would like to
thank the other members of this collaboration for discussions. P.S. acknowledges support 
by BMBF, the DFG Cluster of Excellence ``Origin and Structure of the Universe", and the TUM Graduate School.
J.B. acknowledges support by HIC for FAIR within the LOEWE program of the State of Hesse and by the DFG 
through grant SFB 1245. F.R.~acknowledges support by
FWF grant P24780-N27 and the DFG via SFB 1225 (ISOQUANT).

\bibliography{references}
\end{document}